\documentclass[prl,letterpaper,aps,10pt,superscriptaddress,twocolumn,floatfix,showpacs]{revtex4-1}
\usepackage{graphicx}
\usepackage{amsmath}
\usepackage{amsfonts}
\usepackage{float}
\usepackage{amssymb}
\usepackage{epsfig}
\usepackage{epstopdf}
\DeclareGraphicsExtensions{.pdf,.eps,.png,.jpg,.mps}
\usepackage[pdftex]{color}
\usepackage{amsmath,graphicx,amssymb,braket,xcolor,subfigure,upgreek}
\usepackage[colorlinks, linkcolor=blue, citecolor=blue, urlcolor=blue, breaklinks=true]{hyperref}
\usepackage{microtype}
\usepackage{bbm}
\usepackage{color}
\usepackage{dsfont}

\newcommand{\calS}{\mathcal{S}}

\usepackage{physics}

\begin{document}

\title{Scaling law for Kasha's rule in photoexcited molecular aggregates}
\author{R. Holzinger}
\affiliation{Institut f\"{u}r Theoretische Physik, Universit\"{a}t Innsbruck, Technikerstrasse 21a, A-6020 Innsbruck, Austria}

\author{N. S. Bassler}
\affiliation{Max Planck Institute for the Science of Light, Staudtstra{\ss}e 2,
D-91058 Erlangen, Germany}
\affiliation{Department of Physics, Friedrich-Alexander-Universit\"{a}t Erlangen-N\"urnberg, Staudtstra{\ss}e 7,
D-91058 Erlangen, Germany}
\author{H. Ritsch}
\affiliation{Institut f\"{u}r Theoretische Physik, Universit\"{a}t Innsbruck, Technikerstrasse 21a, A-6020 Innsbruck, Austria}
\author{C. Genes}
\affiliation{Max Planck Institute for the Science of Light, Staudtstra{\ss}e 2,
D-91058 Erlangen, Germany}
\affiliation{Department of Physics, Friedrich-Alexander-Universit\"{a}t Erlangen-N\"urnberg, Staudtstra{\ss}e 7,
D-91058 Erlangen, Germany}
\date{\today}

\begin{abstract}

We study the photophysics of molecular aggregates from a quantum optics perspective, with emphasis on deriving scaling laws for the fast non-radiative relaxation of collective electronic excitations, referred to as Kasha’s rule. Aggregates exhibit an energetically broad manifold of collective states with delocalized electronic excitations originating from near field dipole-dipole exchanges between neighboring monomers. Photo-excitation at optical wavelengths, much larger than the monomer-monomer average separation, addresses almost exclusively symmetric collective states, which for an arrangement known as H-aggregate, show an upward hypsochromic shift. The extremely fast subsequent non-radiative relaxation via intramolecular vibrational modes populates lower energy, subradiant states, resulting in an effective inhibition of fluorescence. Our analytical treatment allows for the derivation of an approximate scaling law of this relaxation process, linear in the number of available low energy vibrational modes and directly proportional to the dipole-dipole interaction strength between neighbouring monomers.
\end{abstract}


\maketitle
Molecular aggregates~\cite{hestand2018expanded,saikin2013photonics,ma2021stacking} are self-ordered arrangements of monomers showing strong collective optical transition dipole strengths. Owing to the dense packing of monomers, with monomer-monomer separations at the order of a nanometer and total aggregate lengths of tens or hundred of nanometers, thus much below an optical wavelength and despite inhomogeneous broadening and separation disorder, they exhibit delocalized excitons~\cite{abramavicius2009coherent}. This allows strong coupling to external light modes, resulting in collectively modified fluorescence rates. Following the discovery of J- and H-aggregates in the $1930$s by Scheibe~\cite{scheibe1937polymerisation} and Jelley~\cite{jelley1936spectral} their standard understanding is based on the original approach introduced by Kasha in the 1960s~\cite{kasha1960energy}. Currently, J-aggregates are widely employed in light-matter coupling experiments aiming at the modification of material properties via the manipulation of the electromagnetic vacuum mode density around electronic resonances~\cite{garcia2021manipulating}. Numerous approaches to understand their aggregation behavior and the subsequent response to external illumination have been taken, as thoroughly reviewed in Ref.~\cite{hestand2018expanded}.\\
\indent There has been recently a new surge in interest to theoretically understand the behavior of organic molecular systems under the action of confined vacuum fields, i.e. within the context of cavity quantum electrodynamics~\cite{berman1994cavity,haroche1989cavity,walther2006cavity,haroche2013exploring}. Among many approaches taken, a purely quantum optics formalism has been introduced, which is based on a simple model for electron-vibron interactions contained within the formalism of the Holstein Hamiltonian and which can incorporate loss of photons or vibrations via open system approach following either a master equation of a quantum Langevin equations approaches~\cite{reitz2019langevin,holzinger2022cooperative}.
\begin{figure}[t]
\includegraphics[width=0.95\columnwidth]{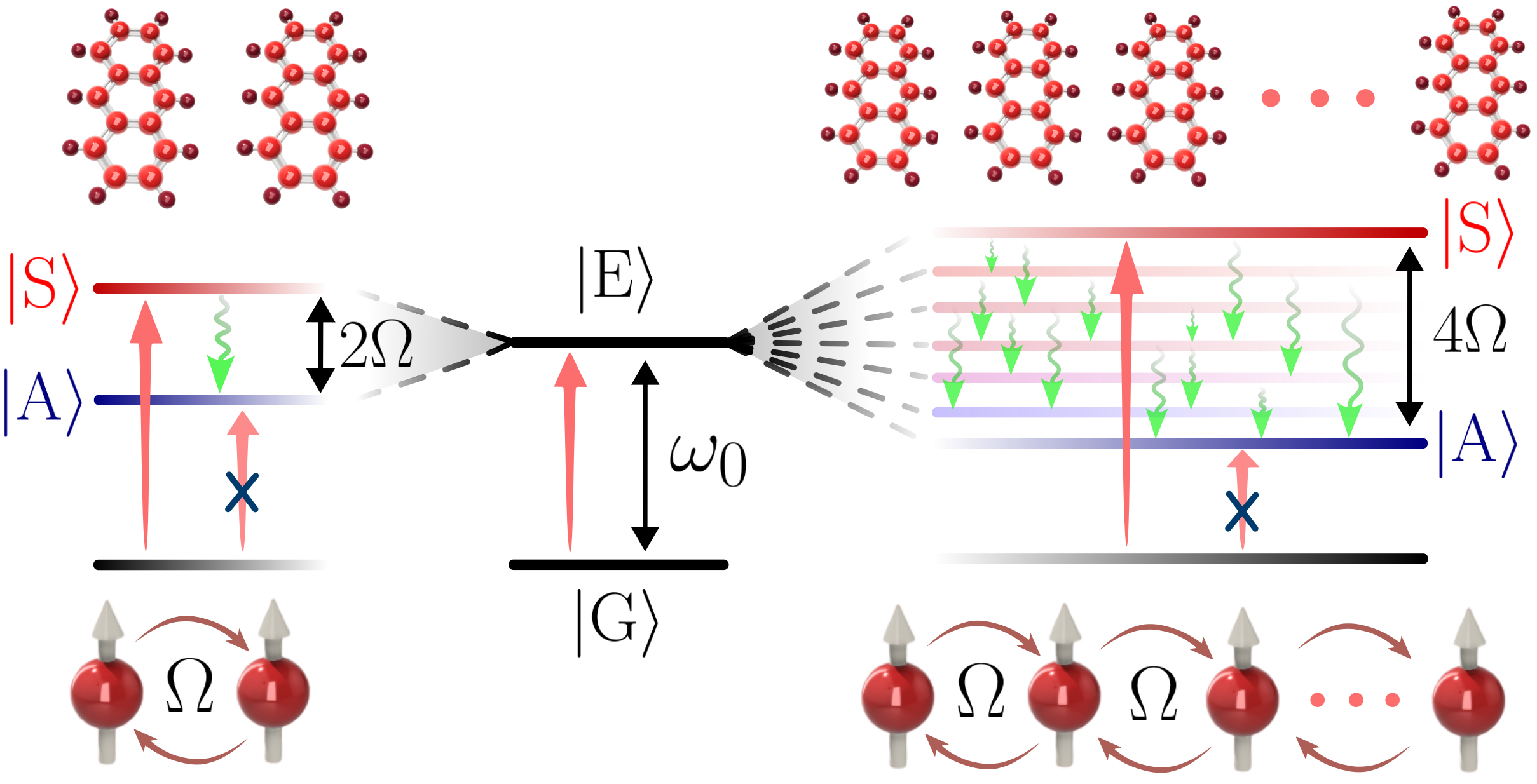}
\caption {Illustration of the Kasha process in an H-aggregate for a dimer (left) and larger aggregate (right). The population migrates from the optically addressed state $\ket{E}$ (superradiant, therefore color coded in red) towards the lower and subradiant collective states $\ket{A}$ (color coded towards blue) via a multitude of vibrational relaxation channels (indicated as green arrows). The energy splitting in the large aggregate is four times the nearest neighbor dipole-dipole coupling strength $\Omega$ and strongly depends on the aggregate separation at the level of nanometers. A resulting effective relaxation rate from the high energy optically addressed state to the bottom of the energy manifold results, on the femtosecond timescale.}
\end{figure}
\indent Within such a quantum optics approach, rate equations for the dynamics of a molecular aggregate upon photo-excitation with light at visible wavelengths $\lambda$ can be derived. As pointed out in Ref.~\cite{hestand2018expanded}, the dynamics of the exciton in the aggregate can be understood in terms of bright (or symmetric) and dark (or asymmetric) states. In the simplest example, a molecular dimer in the side-by-side H-aggregate configuration (see Fig.\ref{fig1}, exhibits a symmetric superposition $\ket{S}=(\ket{eg}+\ket{ge})/\sqrt{2}$ sitting at energy $\omega_0+\Omega$. Here $\ket{eg}$ denotes that the first monomer is in the excited electronic state while the second one is in the ground state and $\Omega$ is the nearest neighbor dipole-dipole interaction scaling with the monomer separation as $d^{-3}$. As $d$ is typically much smaller than $\lambda$, the asymmetric state, at energy $\omega_0-\Omega$, is decoupled from external light and instead is reached indirectly from the symmetric state $\ket{S}$ via redistribution of energy into molecular vibrations, consistent with the standard understanding of a Kasha process. This can be easily generalized to a mesoscopic sized aggregate, where an additional large number of dark states fill the $4\Omega$ frequency split between the photoexcited symmetric state and the lowest energy state. The Kasha incoherent loss of energy is mediated by the multitude of vibrational relaxation pathways. For a molecule with $\mathcal{N}$ atoms a large number of the order $3\mathcal{N}-5$ vibrational modes are available; some of them, up to some index $n_\text{max}$, find themselves within the interval $4\Omega$ thus providing resonant transfer between the set of collective electronic states. 

\indent We show that an approach based on rate equations allows for an approximate expression for the rate of a Kasha process as
\begin{equation}
\label{result}
\kappa_\mathcal{S} \approx \frac{4 s \Omega}{3}\frac{(n_\text{max}+1)(2n_\text{max}+1)}{n_\text{max}}.
\end{equation}
The simplicity of the result lies in the small number of parameters involved: the average estimated Huang-Rhys factor $s=\sum_{j=1}^{n_\text{max}}s_j/{n_\text{max}}$, the monomer separation $d$ (appearing via $\Omega$) and the estimated number of intramolecular vibrations $n_\text{max}$ of relevant frequencies. As an example, we consider the aqueous dye molecule Cresyl Violet which aggregates in an H-configuration, featuring more then $30$ vibrational modes in the range $\sim 5-50$ THz with a zero-phonon transition wavelength $\lambda_0=590$ nm and a monomer fluorescence linewidth $\gamma_0 = 454$ MHz~\cite{jafari2007electronic}. The nearest neighbour dipole-dipole coupling is estimated at around $\Omega=23$ THz and the vibrational spectrum is extracted from Ref.~\cite{barclay2022characterizing}. We estimate around $n_\mathrm{max}=12$ vibrational modes to have a non-negligible Huang-Rhys factor larger than $0.01$ and to fall into the range of the electronic energy band $[0,4\Omega]$; this leads to an average estimated Huang-Rhys factor $s = 0.09$. The estimated timescale for the Kasha process is then around $\kappa_\mathcal{S}^{-1} \approx 13$ fs. Similar estimates for Rhodamine $800$~\cite{christensson2010electronic} and Bacteriochlorophyll a (BChla)~\cite{ratsep2011demonstration} give values for the Kasha process timescale of order $30$ fs and $34$ fs, respectively.\\

\begin{figure}[t]
\includegraphics[width=0.95\columnwidth]{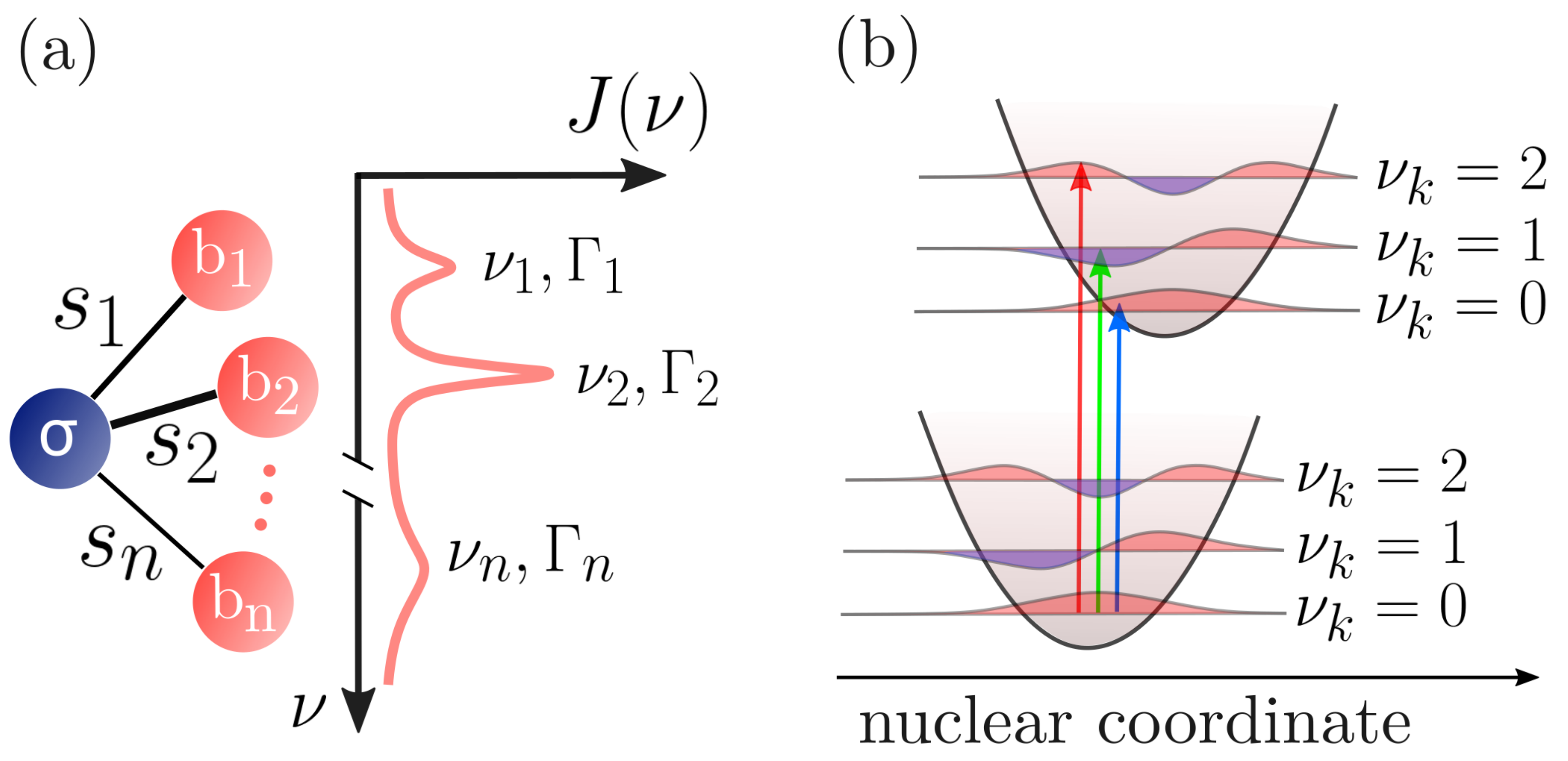}
\caption {(a) Diagramatic description of electron-vibron interactions within each monomer. A single optically addressable electronic transition operator $\sigma$ is coupled to $n$ vibrational mode operators $b_m$ with frequencies $\nu_m$, linewidths $\Gamma_m$ and Huang-Rhys factors $s_m$, where $m$ runs from $1$ to $n$. The vibrational spectral density $J(\nu)$ has a number vibrational frequencies $\nu_m$ and associated relaxation rates $\Gamma_m$ which dictate their linewidths. (b) Ground and excited electronic pootential energy landscapes along a single nuclear coordinate under the harmonic assumption. The vertical lines illustrate the Franck-Condon principle where substantial transition rates only are possible when a match between the symmetry of the nuclear wavepacket in the ground and excited electronic states.}
\label{fig1}
\end{figure}

\noindent \textbf{Photophysics of molecular aggregates} - The minimal model we consider involves a one dimensional molecular chain of identical monomers with a typical inter-monomer distance $d$ in the nm range. As photo-excitation with wavelength $2\pi/k_0$ (with $k_0$ the wavevector) is in the $\mu$m range, the condition $k_0 d\ll 1$ is fulfilled meaning that the chain is uniformly illuminated. Each monomer is assumed to undergo a single electronic transition which in turn is coupled to a $n$ vibrational modes (see Fig.~\ref{fig1}). Each vibrational mode has a frequency $\nu_m$ and relaxation rate $\Gamma_m$, where $m$ runs from $1$ to $n$. The monomer can undergo spontaneous emission at rate $\gamma_0$, owing to the coupling to the electromagnetic environment. For each monomer $j$, the electronic transition is at frequency splitting $\omega_0$ ($\hbar=1$) and is described by the collapse operator $\sigma_j=\ket{g}_j\bra{e}_j$. The vibrational degrees of freedom are described by bosonic operators $b_{jm}$ satisfying the commutation relations $\left[b_{jm},b_{j'm'}^\dagger\right]=\delta_{jj'}\delta_{mm'}$. The vibronic couplings are illustrated in Fig.~\ref{fig1}(a) as links between the electronic and vibration operators with magnitude characterized by the Huang-Rhys factors $s_m$. The electronic and vibrational degrees of freedom are subject to loss quantified by the spontaneous emission rate $\gamma_0$ and by the vibrational relaxation rates $\Gamma_m$, respectively. The Hamiltonian for all $\mathcal{N}$ monomers is obtained as a sum over each particle's Hamiltonian
\begin{equation}
\label{holstein}
h^{(j)}=\bar{\omega}_0\sigma^\dagger_j\sigma_j+\sum_{m=1}^{n}\nu_m \Big( b_{jm}^\dagger b_{jm}- \sqrt{s_m}  \sigma_j^\dagger\sigma_j (b_{jm}^\dagger+b_{jm}) \Big).
\end{equation}
Notice that the bare frequency is Stokes shifted $\bar{\omega}_0=\omega_0+\sum_{m}s_m \nu_m$. This shift will later be eliminated after a polaron transformation (see SI for more details). The model and its validity and relevance is extensively discussed in Ref.~\cite{reitz2019langevin}.

Dipole-dipole exchanges at rates $\Omega_{jj'}$ have a strong imprint at nm distances, owing to their scaling with the inverse cube of the particle separation $|\vec{r}_j-\vec{r}_{j'}|^{-3}$ in the near field region~\cite{reitz2022cooperative}. This can be listed in the Hamiltonian as $\mathcal{H}_\text{d-d}=\textstyle \sum_{j\neq j'}\Omega_{jj'}\sigma^\dagger_j\sigma_{j'}$
and describes an excitation transfer between pairs of monomers via a virtual photon exchange. The coherent exchange is mediated by the dipole-dipole frequency shifts $\Omega_{jj'}$, which in units of the optical emission rate $\gamma_0$ are given by $\Omega_{jj'}/\gamma_0 = -3\pi/k_0 \ \vec{\mu}^* \cdot \mathrm{Re} \ \pmb{G}(\vec{r}_{j},\vec{r}_{j'},\omega_0) \cdot \vec{{\mu}}$, namely, proportional to the real part of the Green's tensor in free space (shown in the Supporting Information). In the following we will consider the particular case of side-by-side arrangement, where all transition dipoles $\vec{\mu}$ are parallel to each other and perpendicular to the chain direction.

\begin{figure}[t]
\includegraphics[width=0.95\columnwidth]{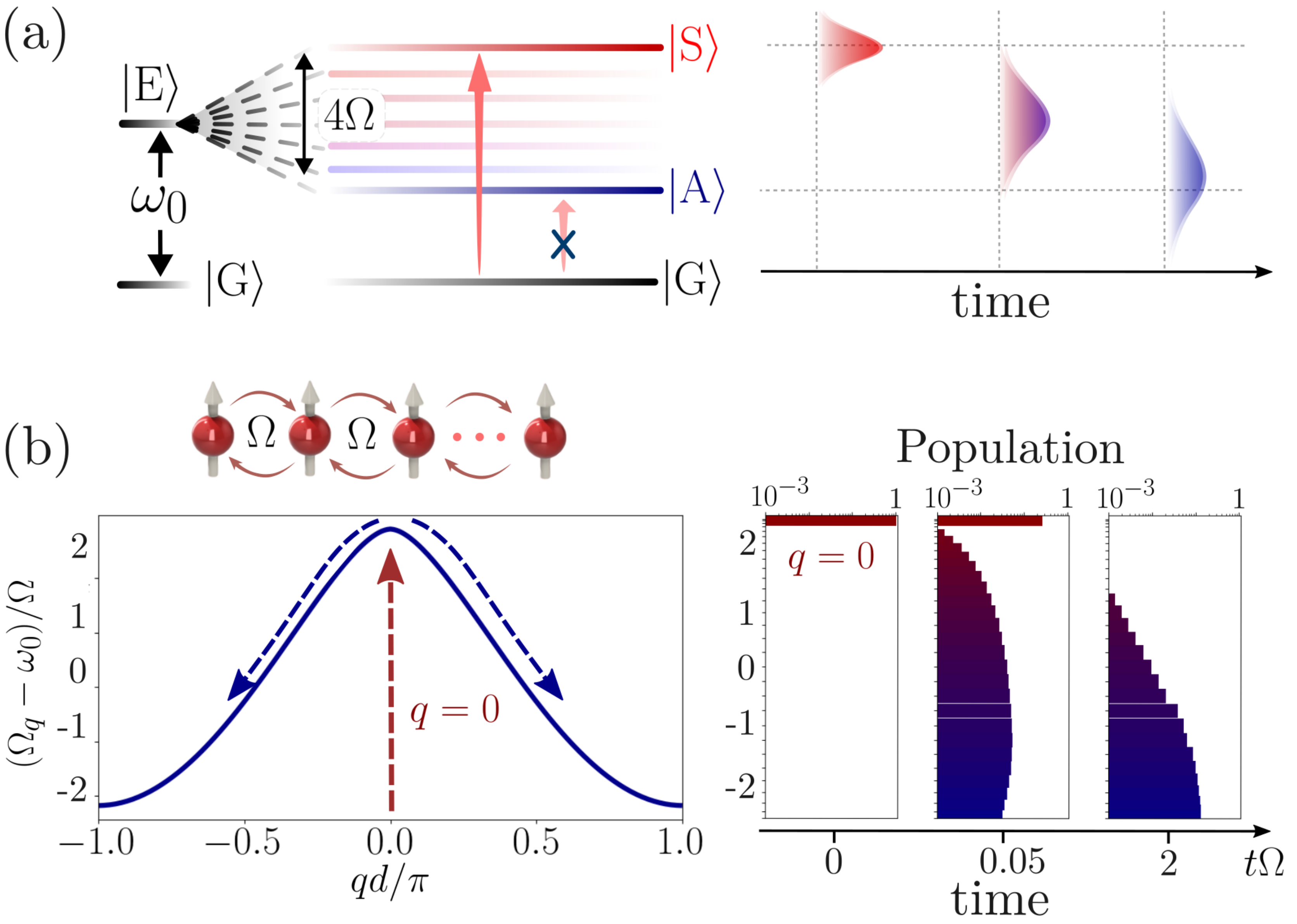}
\caption {Dynamics of the collective state populations upon photoexcitation of the symmetric state in H-aggregates. (a) Illustration of the population migration in time from the optically excited state (superradiant and therefore color coded in red) towards the lower and subradiant collective states (color coded towards blue) via a multitude of vibrational relaxation channels. (b) Exact numerical time dynamics for a chain of $\mathcal{N}=100$ monomers. In such a mesoscopic limit, the analysis can be performed on a band diagram where the symmetric state, photoexcitable, sits at $q=0$. The numerical simulation shows the time dynamics of populations on a time scale normalized to $\Omega^{-1}$.}
\label{fig2}
\end{figure}

The model can be better tackled in a collective basis~\cite{holzinger2022cooperative} resulting from the diagonalization of the first excitation subspace including dipole-dipole interactions. This can be done either in the discrete space where states are indexed from $1$ to $\mathcal{N}$ (see Fig.~\ref{fig2}(a)) or alternatively, with the quasi-momentum index $q = 2 \pi k/(\mathcal{N}d)$ obtained by a re-scaling of the index of the mode $k = \pm 1,..., \pm(\mathcal{N}-1)/2$ (see Fig.~\ref{fig2}(b)). This is based on the observation that, for aggregates with $\mathcal{N}$ of the order of hundred monomers, the mesoscopic limit can be assumed such that the system can be considered translationally invariant and periodic boundary conditions can be invoked. A single symmetric mode can be distinguished with state vector obtained by the application of the symmetric operator $\mathcal{S}^\dagger=\sum_j \sigma_j^\dagger/\sqrt{\mathcal{N}}$ to the collective ground state $\ket{G}$. To a good approximation the system can be considered to be in the Dicke limit where a single superradiant emission rate roughly estimated by $\gamma_\mathcal{S} = \mathcal{N}\gamma_0$ characterizes this 'bright' state. In addition, the other orthogonal $\mathcal{N}-1$ asymmetric states are obtained via the application of asymmetric operators $\mathcal{A}_q = \sum_{j=1}^{\mathcal{N}} {e^{ i  q j d }}\sigma_j/\sqrt{\mathcal{N}}$  to $\ket{G}$. These states are non-radiative at such deep subwavelength molecular separations and we dub them therefore as 'dark'.

The collective excitations are eigenstates of the dipole-dipole interaction Hamiltonian $\mathcal{H}_\mathrm{d-d} \mathcal{A}_q | g \rangle^{\otimes \mathcal{N}} = \Omega_q \mathcal{A}_q | g \rangle^{\otimes \mathcal{N}}$
where by definition we fix the symmetric shift $\Omega_\mathcal{S} \equiv \Omega_{q=0}$. With periodic boundary conditions imposed, in the mesoscopic limit, one can derive the collective shifts as $\Omega_q = 2 \sum_j  \Omega_{1j} \cos( q(j-1)d)$~\cite{asenjo2017exponential,holzinger2022cooperative}.
Further simplifications occur by considering the nearest neighbour approximation and the collective eigenenergies become $\Omega_q = 2 \Omega \cos(qd)$, where the nearest-neighbour coupling is simply denoted by $\Omega \equiv \Omega_{12}$. Closely following the procedure introduced in Ref.~\cite{holzinger2022cooperative}, one can analyze the coupling between states of different symmetries via electron-vibron couplings by an additional transformation to a collective basis for the vibrational degrees of freedom as well. This is done by introducing collective vibrational modes $Q_q^{(m)} = \sum_{j=1}^\mathcal{N}  (b_{m,j} + b_{m,j}^\dagger)e^{ i q j d }/\sqrt{\mathcal{N}},$
with the momentum quadratures satisfying $[Q^{(m)}_q,P^{(m)}_q]=i$ and $m$ labels the vibrational mode running from $1$ to $n$.

\noindent \textbf{Dynamics of the Kasha process in H-aggregates} - Owing to small dimension of a molecular aggregate compared to an optical wavelength, under photo-excitation only the symmetric collective mode is directly accessible. This immediately leads to a rescaling of the Rabi driving by a factor of $\sqrt{\mathcal{N}}$: this can be seen equivalently as an increase of the oscillator strength by $\sqrt{\mathcal{N}}$, thus rendering aggregates of any kind as good candidates for strong light-matter coupling. In addition, the particularity of the subsequent aggregate electronic dynamics lies within the shape of the energy band. For example, J-aggregates present an energy band where the symmetric mode lies at the bottom of the band thus leading to a bathochromic frequency shift (to the left of the bare monomer frequency) and subsequently shows not only an enhanced absorption cross section but also enhanced fluorescence at a superradiant rate. In contrast, H-aggregates have the symmetric state located at the top of the energy band corresponding to a hypsochromic shift. Most importantly, quick dynamics follows the optical excitation involving the relaxation of the collective state towards low energy dark states. This dynamics is illustrated in Fig.~\ref{fig2} in both the discrete and continuous cases. In Fig.~\ref{fig2} a numerical simulation for $\mathcal{N}=100$ at a inter-molecular spacing $k_0 d = 0.0126$. We have considered a large n spanning the range of frequencies from $0$ to $4\Omega$, with $\Gamma_m=\nu_m/10$ and identical $s_m = 0.01$.
Such time dynamics can be analytically tackled at the level of rate equations. We largely follow the derivation in Ref.~\cite{holzinger2022cooperative} which we generalize here to incorporate the crucial aspect that many vibrational modes have to be taken into account. Under the assumption that the vibrational relaxation rates are fast compared to the coherent couplings and radiative loss rates, a set of rate equations for the populations of the symmetric state $p_\mathcal{S}=\langle \mathcal{S}^\dagger \mathcal{S} \rangle$ and all dark states $p_q=\langle \mathcal{A}_q^\dagger \mathcal{A}_q \rangle$ can be derived (see SI). With the definitions $\kappa_{\mathcal{S}}$ - total loss rate of the symmetric state, $\kappa_{q}$ - loss rate for dark state $q$, $\kappa_{q \rightarrow \mathcal{S}}$ - incoherent repopulation rate from the dark to the bright state and $\kappa_{q' \rightarrow q}$ - incoherent rate for redistribution of energy within the dark state manifold, one can write
\begin{subequations}
\begin{align}
\label{populations-sym}
\dot{p}_\mathcal{S} &= -(\gamma_\mathcal{S}+\kappa_{\mathcal{S}}) p_\mathcal{S} + \sum_{q\neq 0} \kappa_{q \rightarrow \mathcal{S}} p_{q}, \\
\dot{p}_{q}&= -\kappa_{q} p_{q} + \sum_{q'\neq q} \kappa_{q' \rightarrow q} p_{q'}.
\end{align}
\end{subequations}

The rate equations show that the symmetric state energy spills into the whole dark state manifold via rate $\kappa_{\mathcal{S}}$ and in addition, higher energy dark states spill into the lower energy ones via $\kappa_{q}$. The quasi-unidirectionality of the process is ensured by the fact that, in this perturbative treatment, the coherent coupling between states is followed by quick vibrational relaxation, making the reverse process, governed by rates $\kappa_{q \rightarrow \mathcal{S}}$ and $\kappa_{q' \rightarrow q}$ from lower energy state to higher ones, very unlikely (as shown in Ref.~\cite{holzinger2022cooperative} for the two monomer case). Mathematically, the condition is $\sqrt{s_m} \nu_m /\sqrt{\mathcal{N}}\ll\Gamma$. Analytically, one can get an expression for the transfer rate between the symmetric mode and any dark mode $q$ mediated by vibrational mode $m$ as
{\begin{equation}
 \label{transfer-rate}
 \kappa^{(m)}_{\mathcal{S} \rightarrow q} = \frac{2s_m \nu_m^2 (\Gamma_{m} +\gamma_{\mathcal{S}})/{\mathcal{N}} }{(\Gamma_{m}+\gamma_\mathcal{S} )^2+ 4(\Omega_\mathcal{S} -\Omega_q-\nu_m)^2}
\end{equation}}
and construct the total rate to all states spanned by the index $q$ by summing $\kappa_{\mathcal{S}}=\sum_{m=1}^{n_\text{max}}\sum_q \kappa^{(m)}_{\mathcal{S} \rightarrow q}$ over all vibrations up to an index $n_\text{max}$ within the frequency interval covered by $4\Omega$ where the electronic collective states are positioned in energy. Moreover, we will consider the standard underdamped harmonic oscillator model for the molecular vibrations, i.e. the dissipation rate is much smaller than the resonance frequency for any mode $\Gamma_{m}\ll \nu_m$. Equivalently, one can state that the quality factor of any vibrational mode is much larger than unity $\nu_m/\Gamma_m\gg 1$.\\
\begin{figure*}[ht!]
\includegraphics[width=0.95\textwidth]{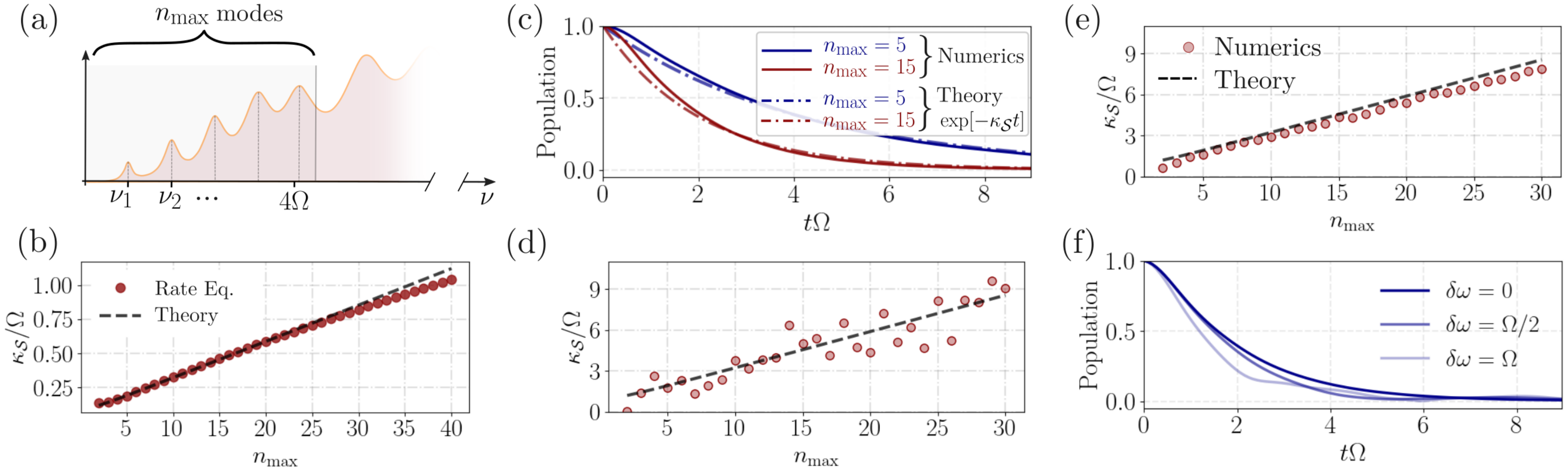}
\caption {(a) Only a limited number of vibrational modes, from $1$ to $n_\mathrm{max}$ can efficiently mediate resonant transfer between the symmetric state and the dark manifold. These modes are within the $4\Omega$ bandwidth. (b) Linear scaling of the transfer rate $\kappa_\mathcal{S}$ as a function of the number of vibrational modes, with the same parameters as in (c) and $\mathcal{N}=20$ molecules at $k_0 d=0.0126$ separation.  The scaling law from Eq.~\eqref{scaling-law} provides a very good fit to the rate equations in Eq.~\eqref{transfer-rate}. (c) Time dynamics of $20$ molecules initialized in the symmetric state. Numerical results in the single excitation manifold show excellent agreement with an exponential decay given by $e^{-\kappa_\mathcal{S}t}$, and governed by the analytical formula in Eq.~\eqref{scaling-law} (Same parameters as in (d)-(e)). (d) Numerical results for $\kappa_\mathcal{S}$ plotted against increasing $n_\text{max}$ for a randomly drawn set of vibrational frequencies $\nu_1,...\nu_{n_\text{max}}$ in the window $0$ to $4\Omega$ and Huang-Rhys factors $s_1,...s_{n_\text{max}}$ in the interval $[0,0.2]$. The fit is performed with the analytical scaling in Eq.~\eqref{scaling-law} for evenly spaced vibrational frequencies and Huang-Rhys factor at the level of the distribution average $s=0.1$. (e) Further comparison of the analytical scaling from Eq.~\eqref{scaling-law} with an average over $200$ random realizations shows almost perfect agreement. (f) Decay of the initial symmetric state population under the influence of random static frequency disorder with fluctuation $\delta \omega$ around $\omega_0$.
Further parameters in all plots: $\mathcal{N}=20$ molecules, $k_0 d = 0.0126$, $\Gamma_m = \nu_m/10$, $\Omega\approx 3.759 \times 10^5 \gamma_0$.}
\label{fig3}
\end{figure*}
In order to further proceed with analytical estimates, let us first make some comments regarding typical timescales. Given that vibrational relaxation is in the order of tens to hundreds of GHz while spontaneous emission is in the range of tens of MHz, a very quick non-radiative path from the symmetric to low energy asymmetric states can be achieved on ps timescales. For monomer separations in the nm range, expected near field shifts $\Omega$ in the range of $1$ THz to tens of THz are expected. This means that only a few, low energy, molecular vibrations can fit in the window of $4\Omega$ (see Fig.~\ref{fig3}(a)) and aid the relaxation process. This allows us to derive an approximate scaling law for $\kappa_\mathcal{S}$ as a function of a given number of vibrational modes $n_\text{max}$ that can efficiently mediate the relaxation of the symmetric state into the dark state manifold. We proceed by first consider a given vibrational mode $m$ and asking for the condition that this mode can transfer excitation from the symmetric mode to any of the dark states. We notice that with the condition that $\Gamma_m\ll\nu_m$ (it is also implied that $\gamma_\mathcal{S}\ll\nu_m$ even for large $\mathcal{N}$) the resonance condition requires that the states to which resonant transfer can take place are only in the vicinity of the mode $q$ fulfilling $\Omega_q=\Omega_\mathcal{S}-\nu_m$. Of course, in this case one can immediately observe that any modes with $\nu_m>4\Omega$ cannot take part in this transfer. Assuming a constant density of all $\mathcal{N}$ collective states spread within the interval $4\Omega$, we can then estimate that a number of approximately $\epsilon_m=\Gamma_m \mathcal{N}/(4\Omega)$ states fall close to the resonance $\Omega_q=\Omega_\mathcal{S}-\nu_m$, i.e. within the linewidth $\Gamma_m$. Summing over all these contributions gives the total rate for all transitions mediated by mode $m$ to states close to $q$. Summing over all possible relaxation paths that participate in the transfer giving thus an estimate for the total rate
\begin{equation}
 \label{scaling-law}
\kappa_\mathcal{S} =\sum_q \kappa^{(m)}_{\mathcal{S} \rightarrow q} \approx \sum_{m=1}^{n_\text{max}} \frac{\Gamma_m}{2\Omega}\frac{s_m \nu_m^2}{(\Gamma_{m}+\gamma_\mathcal{S})}.
\end{equation}
In order to estimate this quantity, knowledge of the particular nature of the monomer's frequencies, Huang-Rhys factors and vibrational relaxation rates is necessary. However, while later we will numerically investigate random distributions of frequencies and Huang-Rhys factors, we aim first at deriving a simple scaling law. To this end we will proceed by making some simplifying assumptions, among which the first is that the vibrational spectrum is equally spaced in the interval from $0$ up to $4\Omega$. We denote the frequency of mode $m$ by $\nu_m=m 4\Omega/n_\text{max}$. Let us also consider that all Huang-Rhys factors are equal to $s$ (later we compare with a randomized distribution with an average $s$). Moreover, we neglect the contribution of $\gamma_\mathcal{S}$ as it is much smaller than $\Gamma_m$ (this should be typically very well fulfilled as $\gamma_0/\Gamma_m$ is expected to be around $10^{-5}$). Summing over all vibrational modes within the interval of $4\Omega$ gives us an approximated scaling law as the expression already introduced in Eq.~\eqref{result}. The result shows independence of the total number of monomers $\mathcal{N}$ and a quasi linear dependence on the total number of available low frequency vibrational modes which can resonantly participate in Kasha's relaxation process from the high energy symmetric state to the bottom of the dark state manifold. Notice that the predicted timescale is dictated by the nearest neighbor dipole-dipole coupling strength $\Omega$ which in turn depends on the inverse cube of the monomer-monomer separation. In a first step, we can estimate that the analytical scaling is in very good agreement with the results of the rate equations, as seen in Fig.~\ref{fig3}(b).\\
\begin{table}[b]
\centering
\begin{tabular}{|c || c c c c c c|}
 \hline
Dye molecule & d(nm) & $|\Vec{\mu}| (\mathrm{D})$ \ & $\gamma_0$(MHz) & $n_\mathrm{max}$ & $ s $ & $\kappa_\mathcal{S}^{-1}$(fs) \\ [0.5ex]
 \hline\hline
 Cresyl Violet &  $\sim$ 2 & 1.8 & 455 & 12 & 0.09 & 13 \\
 \hline
 Rhodamine 800 & $\sim$ 2 & 1.8 & 525 & 16 & 0.067 & 30 \\
 \hline
 BChl a & $\sim$ 1 & 6.4 & 335 & 25 & 0.07 & 34 \\
 [0.25ex]
 \hline
\end{tabular}
\caption{Estimates of the timescale $\kappa_\mathcal{S}^{-1}$ (femtoseconds) of the Kasha process according to Eq.~\eqref{result}. The specific dyes considered are Cresyl Violet (CV) with $\Omega_0 \approx 23$ THz~\cite{jafari2007electronic,barclay2022characterizing}, Rhodamine 800 with $\Omega_0 \approx 8$ THz~\cite{christensson2010electronic} and Bacteriochlorophyll a (BChla) with $\Omega_0 \approx 6$ THz, measured at room temperature~\cite{ratsep2011demonstration}.}
\label{table:1}
\end{table}
\indent The validity of the analytically obtained result can be easily checked against numerical simulations. In a first step, we compare the numerical results with the analytical scaling of Eq.~\eqref{result}, for an equidistant vibrational spectrum, in Fig.~\ref{fig2}(c). A very good agreement is obtained showing that the time evolution of the symmetric state is well reproduced by an exponential decay following $e^{-\kappa_\mathcal{S}t}$. In the next step we pick a set of randomly drawn vibrational frequencies $\nu_1,...\nu_{n_\text{max}}$ in the window $0$ to $4\Omega$ and Huang-Rhys factors $s_1,...s_{n_\text{max}}$ in the interval $[0,0.2]$ (with an average $s=0.1$). The results plotted in Fig.~\ref{fig3}(a) show that a good fit is obtained with the simplified result of Eq.~\eqref{result} which assumes evenly spaced vibrational frequencies and a Huang-Rhys factor at the level of the distribution average $s=0.1$.  Furthermore, an average of the numerical results over $200$ random realizations predicts an excellent agreement to the linear fit predicted by Eq.~\eqref{result}.\\
\indent As aggregates are immersed in solvents and usually under room temperature conditions, they are expected to present a large inhomogeneous broadening, at the level of THz. Moreover, monomers are not placed in an exactly equidistant chain resulting in further shifts in molecular energies. We consider a distribution of the $\mathcal{N}$ monomer frequencies around $\omega_0$ such that the frequency of each monomer becomes $\omega_0 + \delta_j$ where $\delta_j$ is randomly drawn from a distribution of width $\delta \omega$. In the collective basis, the symmetric state couples to any asymmetric state and acquires a shift as well~\cite{sommer2021molecular}. The coupling of $\mathcal{S}$ to a state $q$ mediated by disorder is simply given by the Fourier transform of the distribution $\delta_q = \frac{1}{\sqrt{\mathcal{N}}}\sum_{j=1}^{\mathcal{N}} {e^{ i  q j d }}\delta_j,$
while the shift of the collective state $\delta_\mathcal{S}=1/\sqrt{\mathcal{N}}\sum_{j=1}^{\mathcal{N}}\delta_j$ is the average of the distribution, thus close to zero. According to Ref.~\cite{sommer2021molecular}, disorder induced couplings introduce an additional loss channel, thus slightly increasing the Kasha rate. This is indeed consistent with numerical simulations shown in Fig.~\ref{fig3}(f) where the dynamics of the symmetric state without disorder and with considerable disorder at $\delta \omega=\Omega$ and $\delta \omega=2\Omega$ are compared. The upshot is that the analytically derived loss rate holds well even for considerable disorder levels.

Finally, let us provide some estimates of the expected timescale for such a Kasha process in the aqueous dye molecule Cresyl Violet, Rhodamine $800$ and Bacteriochlorophyll a (BChla). The results are listed in Table.~\ref{table:1} with parameters estimated from Refs.~\cite{jafari2007electronic,barclay2022characterizing,christensson2010electronic,ratsep2011demonstration}. The upshot is that analytically estimated values for the Kasha timescale run in the tens of femtoseconds.\\

\noindent \textbf{Conclusions and outlook} - In conclusion, we have investigated the time dynamics following the photoexcitation of molecular aggregates in the side-by-side configuration. These are a perfect showcase of cooperative phenomena as their photophysics is naturally characterized by coherent and incoherent effects brought on by the positioning of individual monomers in the near field of each other. Effects widely explored in quantum optics, such as Dicke superradiance and subradiance, naturally occur in the theoretical description of such compounds, albeit in the presence of more complex, additional interactions between electrons and a vast number of molecular vibrations.
We have provided a theoretical approach to the dynamics of collective electronic states making the connection, at the analytical and numerical level, with the physical mechanism introduced long ago by Kasha, stating that photon emission (fluorescence or phosphorescence) occurs in appreciable yield only from the lowest excited state of a given multiplicity. Our analytical conclusions predict that the Kasha loss rate from the symmetric, high energy, optically addressable state to the lower energy states of an H-aggregate is roughly independent of the number of monomers but strongly dependent on the number of low energy vibrational modes which can be excited and then dissipate the accumulated energy afterwards. Further investigations will focus on aspects such as the role of quantum coherence in such systems. An important direction is the application of the methods presented in this manuscript to photosynthetic systems under various conditions of illumination, ranging from spatially and time coherent laser light to spatially and time incoherent light sources.\\

\section {Methods}

Numerical simulations are performed in the single excitation subspace. We follow the time evolution of the system assuming unit population of the symmetric state at the initial time $t=0$. As $\Gamma_m$ is larger than the coherent rates, any excitation of a vibration is followed by quick relaxation, justifying the assumption that any double excitation can be neglected. The basis set is picked as a tensor product $\ket{j}\otimes \ket{j'}^{(m)}$ where by definition $\ket{j}=\ket{g,g,...e_j,...}$ - only emitter $j$ excited electronically  and $\ket{j'}=\ket{0,0,...1_{j'},...}^{(m)}$ - only mode $m$ in emitter $j'$ has one vibrational excitation.
Instead of solving the master equation directly, one can use the quantum jump formalism to evaluate single stochastic quantum trajectories using the Monte Carlo wave function method (MCWF). The advantage is that instead of describing the state of the quantum system by a density matrix of size $\mathcal{N}^4 \times n^2$ the stochastic method only requires state vectors of size $\mathcal{N}^2 \times n$. This is somewhat counteracted by the stochastic nature of the formalism which makes it necessary to repeat the simulation until the wanted accuracy is reached. However, in most scenarios, especially for higher dimensional quantum systems, the necessary number of repetitions is much smaller than the system size $\mathcal{N}^2 \times n$ and therefore using the MCWF method is advantageous (see SI). For the simulation we write
\begin{align}
|\Psi \rangle = \sum_{j,j'=1}^{\mathcal{N}} \sum_{m=1}^n  \alpha_{jj'}^{(m)} |g,g,...e_j,...\rangle \otimes  |0,0,...1_{j'},...\rangle^{(m)},
\end{align}
with coefficients $\alpha_{jj'}^{(m)}$ and where the first part refers to the electronic excitation of molecule $j$ and the second part to the excitation of the $m$-th vibrational mode of molecule $j'$.





\section {Acknowledgments} 
We acknowledge financial support from the Max Planck Society and the Deutsche Forschungsgemeinschaft (DFG, German Research Foundation) -- Project-ID 429529648 -- TRR 306 QuCoLiMa
(``Quantum Cooperativity of Light and Matter''). R.~H. acknowledges funding from the Austrian Science Fund
(FWF) doctoral college DK-ALM W1259-N27. We acknowledge fruitful discussions with Michael Reitz and Johannes Feist. The numerical simulations were performed with the open-source framework QuantumOptics.jl~\cite{kramer2018quantum}.

\bibliography{RefsKasha}

\onecolumngrid

\appendix


\section{Vibronic coupling}
\label{AppendixA}

Let us justify the form of the Holstein Hamiltonian in Eq.~\eqref{holstein} by following a first-principle derivation for a single nuclear coordinate $R$ of effective mass $\mu$. We assume that, along the nuclear coordinate, the equilibria for ground (coordinate $R_g$ , state vector $|g\rangle$) and excited (coordinate $R_e$ and state vector $|e\rangle$) electronic orbitals are different.
Assuming equilibrium positions $R_g$ and $R_e$ for the potential surfaces of electronic ground and excited states, one can write the total molecular Hamiltonian describing both electronic and vibrational dynamics as
\begin{align}
\label{molham}
\mathcal{H}_\text{mol}=\left[\omega_0+\frac{\hat{P}^2}{2\mu}+\frac{1}{2}\mu\nu^2\left(\hat{R}-R_e\right)^2\right]\sigma^\dagger\sigma+ \left[\frac{\hat{P}^2}{2\mu}+\frac{1}{2}\mu\nu^2\left(\hat{R}-R_g\right)^2\right]\sigma\sigma^\dagger,
\end{align}
where $\mu$ is the reduced mass of the vibrational mode and $\sigma = |g \rangle \langle e|$. The kinetic and potential energies are written in terms of the position $\hat{Q}$ and momentum operator $\hat{P}$ describing the nuclear coordinate under consideration, with commutation $[\hat{Q},\hat{P}]=i$. Introducing oscillations around the equilibria $\hat{Q}=\hat{R}-R_{\text{g}}$ and subsequently $\hat{R}-R_{\text{e}}=\hat{Q}+R_{\text{g}}-R_{\text{e}}=:\hat{Q}-R_{\text{ge}}$ we obtain
\begin{equation}
\mathcal{H}_\text{mol}=\frac{\hat{P}^2}{2\mu}+\frac{1}{2}\mu\nu^2 \hat{Q}^2 +\omega_0\sigma^\dagger\sigma-\mu\nu^2 \hat{Q} R_{\text{ge}}\sigma^\dagger\sigma+\frac{1}{2}\mu\nu^2 R_{\text{ge}}^2 \sigma^\dagger\sigma.
\end{equation}
 We can now rewrite the momentum and position operators in terms of bosonic operators $\hat{Q}=q_{\text{zpm}}(b^\dagger+b)$, $\hat{P}=ip_{\text{zpm}}(b^\dagger-b)$. The bosonic operators satisfy the usual commutation relation $[{b},{b^\dagger}]=1$ and the zero-point motion displacement and momentum are defined as $q_{\text{zpm}}=1/\sqrt{2\mu\nu}$ and $p_{\text{zpm}}=\sqrt{\mu\nu/2}$. Reexpressing the terms above yields the Holstein Hamiltonian 
\begin{align}
\label{holsteinhamiltonian}
\mathcal{H}_\text{mol}=(\omega_0+s \nu){\sigma}^\dagger{\sigma}+\nu {b}^\dagger{b}- \sqrt{s}\nu({b}^\dagger+{b}){\sigma}^\dagger{\sigma}.
\end{align}
The dimensionless vibronic coupling strength $s$ is given by $\sqrt{s}=\mu\nu R_{\text{ge}}q_{\text{zpm}}$ ($s$ is called the Huang-Rhys factor and is typically on the order of $\sim 0.01-1$).

\section{Radiative and vibrational loss}
\label{AppendixB}
In a master equation formulation for the system density operator $\rho$ written as $\partial_t \rho=i[\rho, \mathcal{H}]+\mathcal{L}[\rho]$ loss can be included via the Lindblad superoperator $\mathcal{L}_\gamma[\rho] = \gamma_{\mathcal{O}}/2\left[ 2\mathcal{O} \rho(t)\mathcal{O}^{\dagger}- \mathcal{O}^{\dagger}\mathcal{O} \rho(t) - \rho(t)\mathcal{O}^{\dagger}\mathcal{O}\right],$
describing decay at generic rate $\gamma_\mathcal{O}$ through a single channel with a generic collapse operator $\mathcal{O}$. For vibrational loss, the collapse rate for each mode $m$ is $\Gamma_m$ and the corresponding collapse operator is $b_{jm}-\sqrt{s_m} \sigma_j^\dagger \sigma_j$. This form for the collapse operator is derived in analogy to the dissipative physics of optomechanical systems in the ultrastrong coupling regime~\cite{dan2015quantum}. The radiative loss is not in diagonal Lindblad form but achieves the following expression $\mathcal{L}_e[\rho] = \sum_{j,j'}\gamma_{jj'}/2\left[ 2\sigma_{j} \rho\sigma^{\dagger}_{j'}- \sigma^{\dagger}_j\sigma_{j'} \rho - \rho\sigma^{\dagger}_j\sigma_{j'}\right]$.
This form can be diagonalized and it shows the emergence of $\mathcal{N}$ independent decay channels, each corresponding to some collective electronic superposition state~\cite{reitz2022cooperative}. At very small separation, deep into the subwavelength regime, the fully symmetric superposition decays at a superradiant rate roughly equal to $\mathcal{N}\gamma_0$ while all other states have vanishingly small decay rates (which we will assume in the following to be exactly zero). This is by no means a limitation of our treatment as one can easily generalize this to the case of non-zero decay rates of the dark manifold~\cite{holzinger2022cooperative,reitz2022cooperative}.\\
\section{Hamiltonian in the collective basis}
\label{AppendixB}
 The Hamiltonian coupling the symmetric state to the dark state manifold is then given by
\begin{align}
  \label{symmetric-coupling}
\mathcal{H}_{\text{int}}^{\mathcal{S}\mathcal{A}}=-\sum_{m=1}^n \sum_{q\neq 0} \frac{\sqrt{s_m} \nu_m}{\sqrt{\mathcal{N}}} \Big(Q_q^{(m)} \mathcal{S}^\dagger \mathcal{A}_q + \mathrm{h.c.} \Big),
\end{align}
via collective vibrations. This coupling is responsible for funneling population into the long lived dark state manifold after the initial driving of the fully symmetric state under uniform illumination. This mechanism is fundamental to understand the dynamics associated with Kasha's rule. In addition, within the dark state manifold an all-to-all coupling Hamiltonian acts with the following form
\begin{align}
  \label{asymmetric-coupling}
\mathcal{H}_{\text{int}}^{\mathcal{A}\mathcal{A}}=- \sum_{m=1}^{n}\sum_{q\neq q'}\frac{\sqrt{s_m}\nu_m}{\sqrt{\mathcal{N}}}\Big(Q_{q-q'}^{(m)}\mathcal{A}_q^\dagger \mathcal{A}_{q'} +\mathrm{h.c.} \Big),
\end{align}
and the sum implies that $q,q' \neq 0$.
This indicates that a redistribution of energy takes place within the whole manifold of dark states. After transforming the system Hamiltonian, the energies of the collective states are shifted by the contribution of the symmetric vibrational mode $-\sum_m \sqrt{s_m} \nu_m Q^{(m)}_{0}/\sqrt{\mathcal{N}}$.
The energy shifts can be removed by the collective polaron transformation $U = \prod_{q} \prod_{m=1}^{n} e^{i \sqrt{s_m}/\sqrt{\mathcal{N}} P_0^{(m)} \mathcal{A}_q^\dagger \mathcal{A}_q}$ which leads to a renormalization of the collective state energies as $\bar{\omega}_q=\omega_0+\sum_m s_m \nu_m/2+\Omega_q$.
\section{Vacuum mediated dipole-dipole coupling rates }
\label{AppendixB}

The vacuum mediated dipole-dipole interactions for an electronic transition at wavelength $\lambda_0$ (corresponding wave vector $k_0=2\pi/\lambda_0$) between identical pairs of emitters separated by $r_{ij}$ is given in terms of the free-space electromagnetic Green's tensor  $ \pmb{G}(\vec{r}_{i}-\vec{r}_{j},\omega_0)  \equiv \pmb{G}(\vec{r}_{ij},\omega_0) $, with $\vec{r}_{ij}=\vec{r}_{i}-\vec{r}_{j}$, which reads

\begin{equation}
\pmb{G}(\vec{r},\omega_0) = \frac{e^{i k_0 r}}{4\pi k_0^2r^3}\Big[(k_0^2 r^2+ i k_0 r -1) \mathbb{I}+(-k_0^2 r^2 -3ik_0 r+3)\frac{\vec{r} \otimes \vec{r}}{r^2} \Big],
\end{equation}
where $r=|\vec{r}|$. The Green's function $\pmb{G}_{\alpha \beta}$ is a tensor quantity, with $\{\alpha,\beta \}= \{x,y,z \}$ which is determined by the polarization direction of the dipoles. In order to obtain the dipole-dipole couplings for H-aggregates we chose linear polarization in z-direction for all molecules, namely we take $\vec{\mu}_z^* \cdot \pmb{G} \cdot \vec{\mu}_z$.

\section{Deriving rate equations}
\label{AppendixC}

Starting from the Holstein Hamiltonian for $\mathcal{N}$ identical molecules with $n$ vibrational modes each, the Heisenberg equations for the collective electronic modes are given by

\begin{subequations}
    \begin{align}
        \dot{\mathcal{S}} &= -i \Big( \Omega_\mathcal{S}- \frac{\gamma_\mathcal{S}}{2} \Big) \mathcal{S} +\frac{i\sqrt{s}_m \nu_m}{\sqrt{\mathcal{N}}} \sum_{m=1}^n \sum_{q} Q_q^{(m)} \mathcal{A}_q  + \mathrm{noise}, \\
      \dot{\mathcal{A}_q} &= -i \Omega_q \mathcal{A}_q  + \sum_{m=1}^n \frac{i\sqrt{s_m} \nu_m}{\sqrt{\mathcal{N}}} \Big( {Q^{(m)}_q}^\dagger \mathcal{S} + \sum_{q'\neq q} Q^{(m)}_{q-q'} \mathcal{A}_{q'} \Big)  + \mathrm{noise}.
    \end{align}
    \end{subequations}

The collective noise terms will be neglected from now on as they do not contribute to the transfer process.

To calculate the transfer rate from the symmetric state to the antisymmetric states we assume some initial population in the symmetric state and no population in the antisymmetric states, additionally we assume that the symmetric state decays independently and formally integrate
\begin{subequations}
\begin{align}
\mathcal{S}(t) &= S(0)e^{-(i\Omega_\calS+\gamma_\calS/2)t}, \\
\mathcal{A}_q(t) &= A_q(0)e^{-i \Omega_q t} + \sum_{m=1}^n \frac{i \sqrt{s_m} \nu_m}{\sqrt{\mathcal{N}}}\int^t_0 dt'e^{-i \Omega_q (t-t')}\Big( Q_q^{(m)}(t')\mathcal{S}(t') + \sum_{q' \neq q} Q^{(m)}_{q-q'}(t')\mathcal{A}_{q'}(t')\Big),
\end{align}
\end{subequations}

and for the expectation value of the populations we get
\begin{subequations}
\begin{align}
\dot{\langle \mathcal{S}^\dagger \mathcal{S} \rangle} &= -\gamma_\calS{\langle \mathcal{S}^\dagger \mathcal{S} \rangle} -\sum_{m=1}^n \sum_q \frac{2 \sqrt{s_m} \nu_m}{\sqrt{\mathcal{N}}}  \, \mathrm{Im}\langle \mathcal{S}^\dagger \mathcal{A}_q Q_q^{(m) }\rangle ,\\
\dot{\langle \mathcal{A}_q^\dagger \mathcal{A}_q \rangle} &=  -\sum_{m=1}^n \frac{2 \sqrt{s_m} \nu_m}{\sqrt{\mathcal{N}}} \, \mathrm{Im} \Big( \langle \mathcal{A}_q^\dagger \mathcal{S} Q_q^{(m)}\rangle + \sum_{q'\neq q} \langle \mathcal{A}_q^\dagger \mathcal{A}_{q'} Q_{q'-q}^{(m)}\rangle \Big).
\end{align}
\end{subequations}

Therefore the terms $-2 \sqrt{s}_m \nu_m /\sqrt{\mathcal{N}} \,\mathrm{Im}\langle \mathcal{A}_q^\dagger \mathcal{S} Q_q^{(m)}\rangle$ will be responsible for population transfer from the symmetric to the antisymmetric state with quasi-momentum $q$ at a rate $\kappa^{(m)}_{\mathcal{S} \rightarrow {q}}$. We can calculate the rates explicitly up to order $\mathcal{O}(s_m \nu_m^2)$ and assuming that correlations between vibronic and electronic operators factorize.

\begin{align}
-2 \sqrt{s}_m \nu_m /\sqrt{\mathcal{N}} \langle \mathcal{A}_q^\dagger \mathcal{S} Q_q^{(m)}\rangle &= -i s_m \nu_m^2 \int_0^t dt' e^{-\Omega_{q}(t-t')} \langle Q_{q}^{(m)}(t')Q_q^{(m)}(t)\rangle \langle \mathcal{S}^\dagger(0) \mathcal{S}(0)\rangle e^{-\epsilon_\calS t'} e^{-\epsilon^*_\calS t} \nonumber \\
 &= -i s_m \nu_m^2 \langle \mathcal{S}^\dagger(0) \mathcal{S}(0)\rangle \frac{e^{-\gamma_\calS t} - e^{-((\Gamma_m+\gamma_\calS)/2+i(\Omega_\calS-\Omega_q-\nu_m))t}}{(\Gamma_m+\gamma_\calS)/2 + i(\Omega_\calS-\Omega_q-\nu_m)},
\end{align}

where we defined $\epsilon_\calS = -(\gamma_\calS/2-i\Omega_\calS)$ and used the fact that different vibrational modes are uncorrelated at all times, i.e. $\langle Q^{(m')}_{q}(t') Q^{(m)}_q(t)\rangle=0$ for $m' \neq m$. The correlations for $Q_q^{(m)}$ are evaluated assuming free evolution of the vibrations (to lowest order) and  zero temperature for the vibrational modes:
\begin{equation}
\langle Q_q^{(m)}(t') Q_q^{(m)}(t) \rangle = \frac{1}{\mathcal{N}}\sum_{j=1}^\mathcal{N} \langle b_{jm}(t')b_{jm}^\dagger(t)\rangle =e^{-(\Gamma_m/2-i\nu_m)(t-t')}.
\end{equation}
The transfer rate can be written as
{\begin{equation}
 \kappa^{(m)}_{\mathcal{S} \rightarrow q} = \frac{2s_m \nu_m^2 (\Gamma_{m} +\gamma_{\mathcal{S}})/{\mathcal{N}} }{(\Gamma_{m}+\gamma_\mathcal{S} )^2+ 4(\Omega_\mathcal{S} -\Omega_q-\nu_m)^2},
\end{equation}}
given fast vibrational relaxation rates $\Gamma_m \gg \gamma_\calS$ compared to the electronic decay rates.

\section{Single excitation subspace}
\label{AppendixD}

The numerical diagonalization and subsequent time dynamics are evaluated in the single-excitation sector for both the electronic and vibrational degrees of freedom. This allows to rewrite the effective Hamiltonian in non-hermitian form as ($\hbar = 1$)
\begin{align}
\mathcal{H}_\mathrm{eff} = \sum_{j=1}^\mathcal{N} \Big( h^{(j)} + \sum_{j' =1}^{\mathcal{N}} \Big(\Omega_{jj'} - i \frac{\gamma_{jj'}}{2} \Big)\sigma^\dagger_j \sigma_{j'} - \frac{i}{2} \sum_{m=1}^{n}  \Gamma_m \mathcal{O}^\dagger_{jm} \mathcal{O}_{jm} \Big),
\end{align}
where $h^{(j)}$ is defined in Eq.~\eqref{holstein} and $\mathcal{O}_{jm} = b_{jm}-\sqrt{s_m}{\sigma}^\dagger_j {\sigma}_j$.
The dynamics of the electron-vibron density matrix $\rho$ can be described by a von Neumann equation of the form
\begin{align}
i \frac{\mathrm{d}}{\mathrm{dt}} \rho(t) = [\mathcal{H}_\mathrm{eff} \rho - \rho \mathcal{H}_\mathrm{eff}],
\end{align}
and the expectation value of observable $\mathcal{O}$ becomes $\bar{\mathcal{O}}=\mathrm{tr}(\rho {\mathcal{O}})$.
However, instead of solving the von Neumann equation directly, one can use the quantum jump formalism to evaluate single stochastic quantum trajectories using the Monte Carlo wave function method (MCWF). For large numbers of trajectories, the statistical average then approximates the result of the Master equation. The huge advantage is that instead of describing the state of the quantum system by a density matrix of size $\mathcal{N}^4 \times n^2$ these trajectories work in terms of state vectors of size $\mathcal{N}^2 \times n$. This is somewhat counteracted by the stochastic nature of the formalism which makes it necessary to repeat the simulation until the wanted accuracy is reached. It turns out, however, that for many cases, especially for high dimensional quantum systems, the necessary number of repetitions is much smaller than the system size $\mathcal{N}^2 \times n$ and therefore using the MCWF method is advantageous.

The system size stems from the fact, that in the single excitation subspace for both electronic and vibronic modes a general state vector can be written as
\begin{align}
|\Psi \rangle = \sum_{j=1}^{\mathcal{N}} \alpha_{jj'}^{(m)} |g,g,...e_j,...\rangle \otimes  \sum_{m=1}^n \sum_{j'=1}^{\mathcal{N}} |0,0,...1_{j'},...\rangle^{(m)},
\end{align}
with coefficients $\alpha_{jj'}^{(m)}$ and where the first part refers to the electronic excitation of molecule $j$ and the second part to the excitation of the $m$-th vibrational mode of molecule $j'$.
Thus, the single excitation assumption substantially reduces the Hilbert space dimension from $2^\mathcal{N} \times n_\mathrm{cut}^{\mathcal{N}} \times n$ to $\mathcal{N}^2 \times n$ (where $n_\mathrm{cut}$ is the cut-off of the Fock space dimension for the vibrational modes), allowing the simulation of mesoscopic numbers of molecules.


\end{document}